**Layer Selection in Feature-Based Losses Affects Image Quality and Microstructural Consistency in Deep Learning Super-Resolution of Brain Diffusion MRI**

David Lohr[1,2,3,*], Rene Werner[1,2,3]

[1]Institue for Applied Medical Informatics, University Medical Center Hamburg-Eppendorf, Hamburg, 20246, Germany

[2]Institute of Computational Neuroscience, University Medical Center Hamburg-Eppendorf, 20246, Hamburg, Germany

[3]Center for Biomedical Artificial Intelligence (bAIome), University Medical Center Hamburg-Eppendorf, 20246, Hamburg, Germany

*Corresponding and lead author: d.lohr@uke.de

Parts of this paper have been presented at the 2025 Annual meeting of the ISMRM.

**Summary**

Clinical application of high-resolution diffusion MRI is hindered by hardware limitations and prohibitive scan times, motivating computational super-resolution. This study investigates the efficacy of a feature-based loss function in preserving diffusion signal consistency in deep learning super-resolution. Using 7T data from the human connectome project to generate pairs of low- and high-resolution diffusion weighted images (DWI), we trained UNets for 2D super-resolution. Ablation and isolation studies evaluated different VGG16-layers for feature-based losses against an image-based $L_1$ baseline. Deeper layers and combinations thereof resulted in grid-like artifacts in super-resolution DWIs, which persisted in diffusion parameters like quantitative and fractional anisotropy. No such artifacts were present when using the shallowest layer. Downstream analysis for this layer showed great consistency with the ground truth, even for 9-fold super-resolution. Image SNR and used VGG16-layer depths modulated artifact appearance and severity, mandating careful selection of contributing layers for application in and beyond diffusion MRI.

**Introduction**

As a non-invasive method for the assessment of tissue microstructure, diffusion MRI has become the tool of choice for research on normal brain development, connectivity as well as disease-related microstructural alterations.[1-4] High fidelity analysis of these aspects that allow the distinction of crossing and kissing pathways, as well as restricted and unrestricted diffusion, however, relies on the acquisition of diffusion-weighted images with various diffusion weightings and directions. The 7T acquisition protocol of the human connectome project (HCP)[5], e.g., contains 15 unweighted images and 64 diffusion directions for the b-values of 1000 s/mm$^2$ and 2000 s/mm$^2$ for 1 mm isotropic slices and is repeated for two opposing phase encoding directions. Given a repetition time of TR=7 s, scan time amounts to ~33 minutes. Such protocols are not applicable in clinical practice due to scan time, hardware, and signal-to-noise ratio (SNR) constraints. Deep learning has increasingly been applied to address these limitations by enhancing spatial[6-8] and/or angular[9-12] resolution via computational super-resolution (SR) models.

While many deep learning models are still trained using classical image quality measures like mean absolute error (MAE), mean squared error (MSE), and peak signal-to-noise ratio (PSNR) as a loss function to guide the learning process, these measures do not optimize outputs for human visual perception and the reconstruction of high frequency features. Structured similarity index measure (SSIM)[13], multi-scale SSIM[14], and feature similarity index measure (FSIM)[15] have been developed to address this issue, but come with their own limitations. This has motivated the so-called perception or feature loss, which was originally designed to enable style transfer and super-resolution for natural images based on perceptual similarity.[16] For feature-based losses, both the ground truth (GT) and the prediction of a model are passed through a pre-trained model. During this pass, feature maps are extracted at individual layers. Perceptual similarity and restoration of high frequency features is then achieved using differences in these feature maps as the loss. While such feature-based losses are already applied in MRI,[17-21] there is limited evidence of its benefit or the driving factors behind it. This study thus aims to systematically assess the effects of a feature-based loss in the context of spatial super-resolution in diffusion MRI, a task which requires robustness to large SNR variations, various image contrasts, and the reconstruction of both low and high frequency features.

We believe diffusion MRI to be suitable for this assessment for the following reasons: 1) diffusion scans inherently contain many images due to the diffusion directions acquired, reducing the number of repeated or additional scans required to train or fine-tune these models in different settings. 2) With the HCP repository, there is high fidelity open-source data available to train and evaluate such models and generate baselines. This facilitates reproducibility and provides access to research groups and sites that do not acquire their own data. 3) Since diffusion information is encoded as signal loss, diffusion data exhibits large SNR variations, providing a larger range of testing conditions than anatomical scans. 4) We further hypothesize that feature-based losses are particularly beneficial for diffusion MRI, since diffusion gradients at various weightings lead to distinct intensity patterns in different regions of the brain which represent various microstructure effects on the diffusion process. If these patterns represent key features of brain diffusion images, losses derived directly from these features are likely to preserve the semantic qualities better than pixel-based losses, enabling indirect physics-based guidance for the learning process. The captured structural and contextual information, such as tissue boundaries and microstructural details, further leads to physically consistent features.

Primary aim of this study was an in-depth analysis of feature-based losses in the context of diffusion MRI. This included systematically evaluating the performance of models to accurately enhance the resolution of diffusion MRI data, when trained with loss components from different isolated layers as well as combining features from multiple layers. Utilizing the different configurations, we identified if and how feature-based losses contribute to preserving diffusion information and high frequency features. Secondary aim was to demonstrate SR for diffusion MRI based on a model with small memory footprint for downstream applications such as deriving microstructural parameters or conducting tractography. Only by moving beyond mere diffusion weighted image (DWI) quality analysis, it is ensured that high-resolution diffusion MRI data can be achieved from low-resolution (LR) scans without compromising microstructural insights.

**Results**

Adding the feature loss via VGG16 layer outputs did not affect training times. Training for one epoch took ~42 min for both the baseline (pixel-wise $L_1$ loss) and feature-based models, resulting in a total training time of ~70h per model. PSNR and loss development on the validation set is depicted in Figure S1.

**Image quality scores**

**Experiment 1 - ablation study**

Figure 1 illustrates image quality scores for the ablation study. Depicted are downsampled central slices ($b_0$, $b_{1000}$, and $b_{2000}$) of the test set, respective model outputs and the GT. Visually, model outputs for $b_0$ images showed hardly any difference in image quality compared to the GT. For the $b_{1000}$ images, the feature-based losses led to distinct grid-like noise in the image. The intensity of that noise pattern appeared to be less distinct the fewer layers were contributing to the loss. Higher b-values seemed to aggravate this effect. When only the shallowest layer was involved no such pattern was visible independent of diffusion weighting and SNR.

All image quality scores confirmed the visual assessment that the prediction quality improves when using shallower layers for the feature-based loss. Compared to bicubic upsampling ($b_0$: 26.5±2 dB) both feature-based models and the $L_1$ baseline significantly ($p \leq 0.005$) improved PSNR. The model using a combination of all layers achieved mean PSNR and SSIM of 30.3±2.3 dB and 0.86±0.04 on $b_0$ images, while the model using solely the shallowest layer (red box) achieved 30.9±2.6 dB and 0.88±0.04, respectively. Higher diffusion weighting in the input images led to lower values in image quality scores. The same was found for very caudal and dorsal slice positions. Dependences of the scores PNSR and SSIM regarding diffusion weighting and slice position are illustrated in Figure S2. For $b_0$ images, differences in PSNR and SSIM between feature-based models and the $L_1$ baseline (31.2±2.3 dB and 0.88±0.04) were non-significant ($p \geq 0.34$, $p \geq 0.18$) as were differences between the feature-based models ($p \geq 0.58$, $p \geq 0.19$).

**Experiment 2 - isolation study**

Image quality scores of the isolation study are depicted in Figure 2. While grid-like noise artifacts were barely perceptible for unweighted ($b_0$) images, they were more pronounced for $b_{1000}$ images. There was no clear correlation between artifact intensity and layer depth. Like in the ablation study, higher b-values seemed to aggravate the artifacts, while none were visible for the shallowest layer (red box) independent of diffusion weighting and SNR.

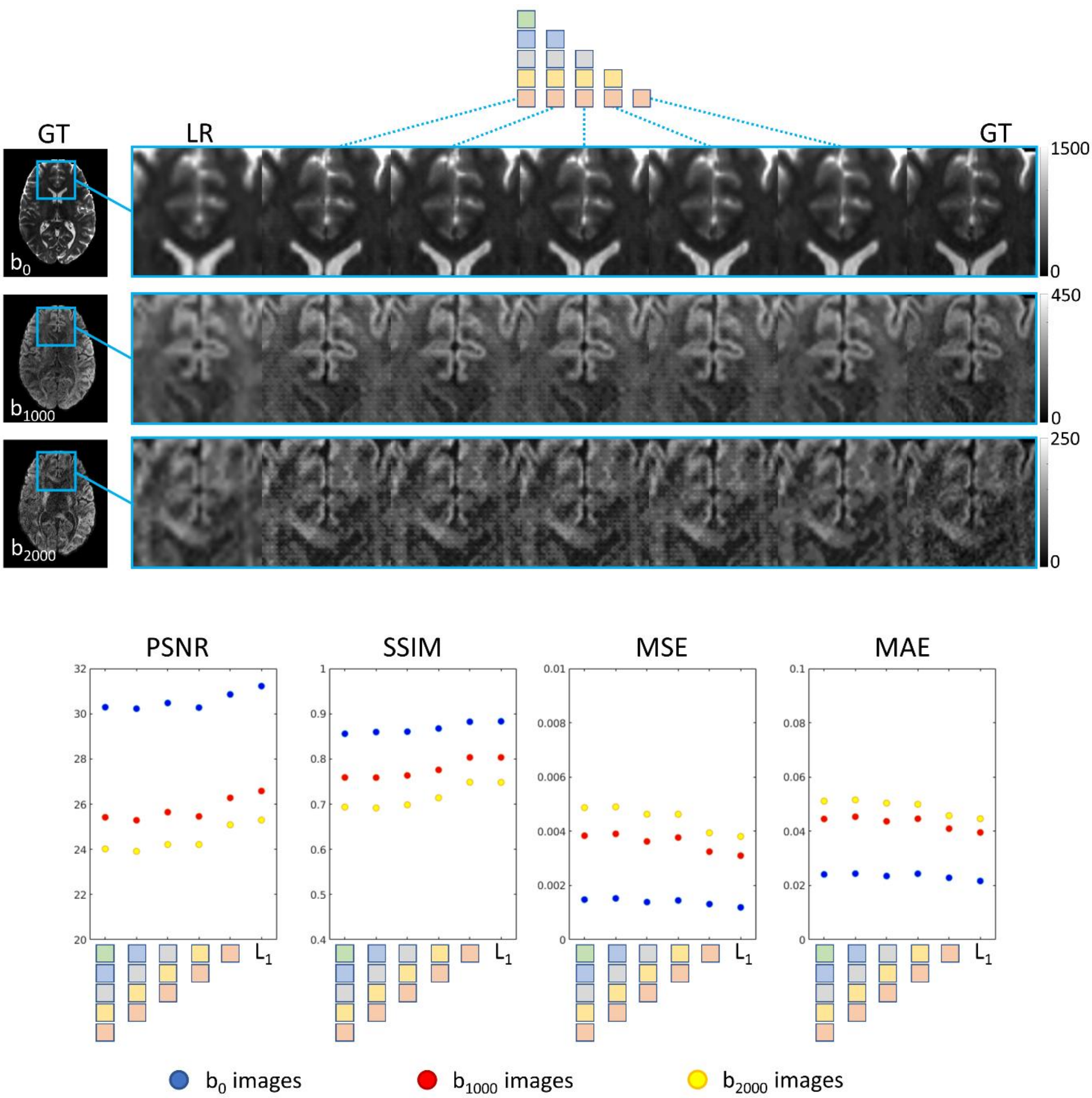

***Figure 1: Ablation study results on the test set.*** *Coloured boxes indicate the layers that contributed to the loss.* ***Top:*** *depicted are a $b_0$, a $b_{1000}$, and a $b_{2000}$ image of a central slices of one subject. Cyan squares indicate crop sections employed for the low-resolution (input), the various model predictions for that input, and the ground truth (GT).* ***Bottom:*** *image quality scores absolute error (MAE), mean squared error (MSE), peak signal-to-noise ratio (PSNR), and SSIM (structural similarity index measure) averaged over all slices and subjects in the test set separated by b-value.*

PSNR values were highest for the model using the central layer (grey box) in the loss (31.1±2.1 dB), whereas SSIM was highest for the shallowest layer (red box, 0.88±0.04) and decreased with layer depth. Compared to bicubic upsampling ($b_0$: 26.5±2 dB) all models again significantly ($p \leq 0.009$) improved PSNR. Higher diffusion weighting as well as very caudal and dorsal slice positions led to lower image quality scores (Figure S3). Differences in PSNR and SSIM between feature-based models and the $L_1$ baseline were non-significant ($p \geq 0.29$, $p \geq 0.10$) as were differences between the feature-based models ($p \geq 0.32$, $p \geq 0.12$). Models

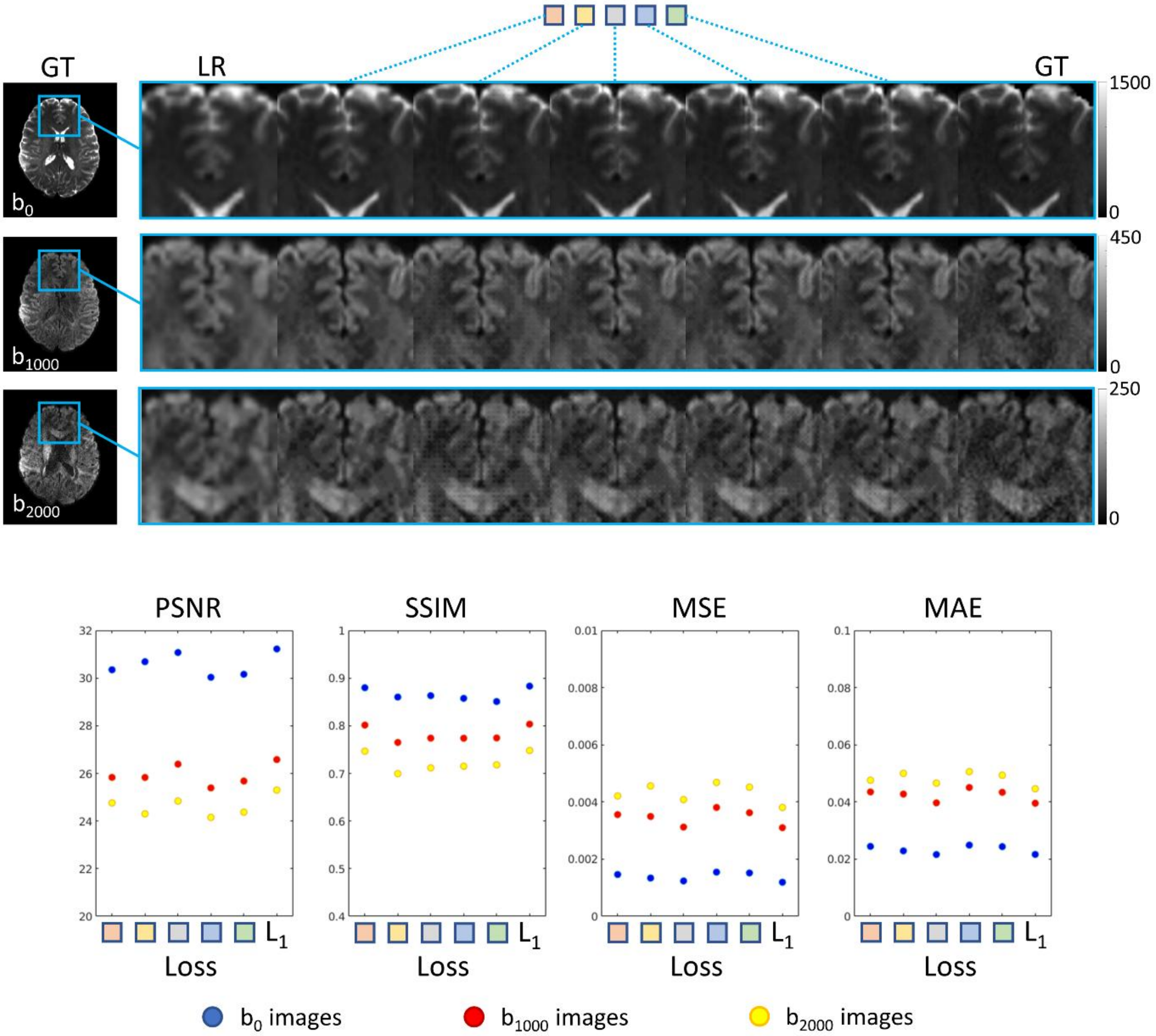

***Figure 2: Isolation study results on the test set.*** *Coloured boxes indicate the layers that contributed to the loss.* ***Top:*** *depicted are a $b_0$, a $b_{1000}$, and a $b_{2000}$ image of a central slices of one volunteer. Cyan squares indicate crop sections employed for the low-resolution (input), the various model predictions for that input, and the ground truth (GT).* ***Bottom:*** *image quality scores absolute error (MAE), mean squared error (MSE), peak signal-to-noise ratio (PSNR), and SSIM (structural similarity index measure) averaged over all slices and subjects in the test set separated by b-value.*

**Quality control of diffusion data**

Values for the diffusion contrast of all models including the baseline (range: 1.158-1.180) deviated non-significantly ($p \geq 0.46$) from the ground truth (1.165±0.032), indicating comparable likelihood to resolve fiber bundle orientations. Correlation values for all feature-based models (range: 0.978-0.980) were comparable to the baseline (0.980±0.003), but significantly higher than for the ground truth (0.972±0.004, p:0.005-0.026) prior to multi-test correction. This indicates higher diffusion coherence within the resolution-enhanced diffusion data. The highest correlation was achieved by the model using solely the first layer (0.980±0.003). Quality scores for individual models are depicted in Figure S4.

**Diffusion parameters**

Figure 3 shows the absolute difference between model-based and ground truth generalized q-sampling imaging (GQI)[22] reconstruction for the diffusion parameters quantitative anisotropy (QA) and restricted diffusion imaging (RDI) in the superior and medial longitudinal fasciculus, the anterior commissure as well as all regions of the HCP842 atlas. In healthy brain tissue higher QA values are often indicative of clear, well-defined structures and are considered a sign of good image quality and meaningful microstructural information. Respective values were derived for all models of the ablation and isolation study. Shallower layers and combinations thereof tended to provide non-significantly ($p \geq 0.38$) higher QA than the ground truth, while simultaneously decreasing variance. With a mean difference of $6.7 \times 10^{-6}$ averaged over all regions of interest (ROIs) of the atlas, highest agreement with the ground truth was found for the model using layer 2 (yellow box).

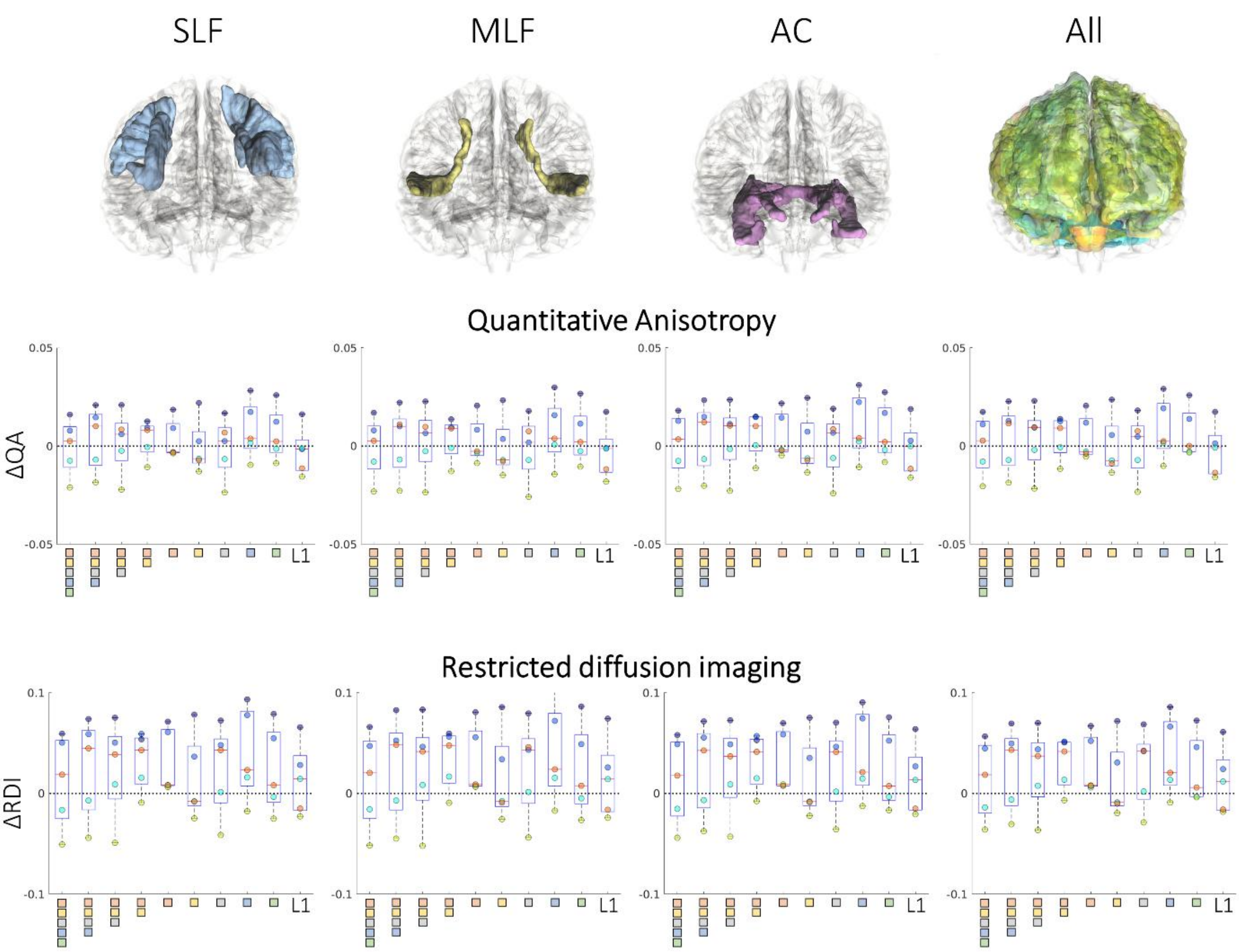

***Figure 3: Region-based diffusion parameters for models of the ablation and isolation study.*** *Regions of interest (ROI) were derived from the HCP842 atlas.* ***Top:*** *ROIs selected to illustrate model performance in various brain regions, namely superior longitudinal fasciculus (SLF), median longitudinal fasciculus (MLF), and anterior commissure (AC). The ROI "All" contains the full 80 ROIs within the atlas.* ***Middle:*** *Quantitative anisotropy (QA) difference between model predictions and ground truth on the test set. Coloured boxes indicate the layers that contributed to the loss. Colour coding for the data points denotes different volunteers of the test set. The dotted line marks where prediction and ground truth would have been equal.* ***Bottom:*** *Difference in restricted diffusion imaging (RDI) between model predictions and ground truth in the test set. Color-coding marks the n=5 test scans. Boxes in boxplots depict 25th to 75th percentiles, medians are marked as horizontal line, and whiskers extend to the outermost data points.*

The baseline model non-significantly ($p=0.72$) underestimated QA, providing a mean difference of $-2.7\times10^{-3}$. For RDI, all model predictions led to higher values compared to the ground truth. Like for QA, shallower layers or combinations therefore provided higher values and lower variance. Averaged over all ROI, highest agreement with the ground truth was found for the model using layer 2 (yellow box, 0.015) and the baseline (0.014). Parameter maps for QA and fractional anisotropy (FA) (Figure S5) reveal that the grid-like artifacts observed in DWIs persisted in derived diffusion parameters, contributing to the higher variances in diffusion parameters for models using deeper layers or combinations thereof.

Trends for the parameters QA and RDI in the individual brain regions superior longitudinal fasciculus (SLF), median longitudinal fasciculus (MLF), and anterior commissure (AC) are very similar and match the average of all ROIs, indicating that different loss terms have a consistent effect on derived diffusion parameters independent of the brain region. Increasingly deep layers and combinations thereof exhibit increased variance, most likely related to the distinct artifact patterns. Distributions for the parameters FA, radial diffusivity (RD), and isotropy (ISO) are showcased in Figure S6. Both diffusion tensor parameters FA and RD tend to be underestimated and variance did not increase for increasingly deep layers or combinations thereof.

**Experiment 3 – resolution study**

Based on the first two experiments, we found the model using solely the first layer of the VGG16 the most consistent with respect to the evaluation criteria applied. Resolution restorations were free of artifacts, diffusion quality scores were high, and GQI diffusion parameters exhibited low variance. Figure 4 depicts tractography based on the ground truth HCP data as well as 4-fold, and 9-fold upsampled data using our baseline model and the model based on the shallowest layer (red box). Models for 9-fold resolution restoration were fine-tuned as described in the method section. Regions for seeding were the AC and the posterior commissure (PC). QA, RDI, and FA for pixels within the AC and PC tracts of the ground truth were 0.27, 0,46, 0,54 and 0.23, 0.46, 0.48, respectively. Values in the 4-fold upsampled data were 0.30, 0.52, 0.52 and 0.26, 0.52, 0.48 for the feature-based model and 0.29, 0.51, 0.52 and 0.26, 0.52, 0.49 for the baseline. Tractography derived from the feature-based model resembled the ground truth more accurately than the baseline, particularly for the PC seed, while areas of thicker tracts exhibited no clear differences. Tractography for the feature-based model resulted in some tracts which were not present in the baseline or the ground truth. Higher QA and RDI compared to the GT were found in MLF and SLF tracts as well for both models. 9-fold upsampling did not result in significant changes in the diffusion metric for the feature-based model and the baseline and similar observations were made in MLF and SLF tracts. Respective tractography based on seeds in the MLF and SLF is shown in Figure S7.

**Discussion**

Using publicly available HCP data, we identified benefits and limitations of feature-based loss functions for SR models in diffusion MRI. We showcase that feature-based losses may cause artifacts and demonstrate that the SNR of the input image and the depth of the used layers modulate the severity of these artifacts. We further show that training a UNet with a small-scale encoder (ResNet18) using a feature-based loss can enable 4-fold to 9-fold SR of diffusion MRI. Most notably, however, our results show that the feature-based loss as introduced by Johnsons et al. for natural images[16] is unsuitable for diffusion MRI without adaptations.

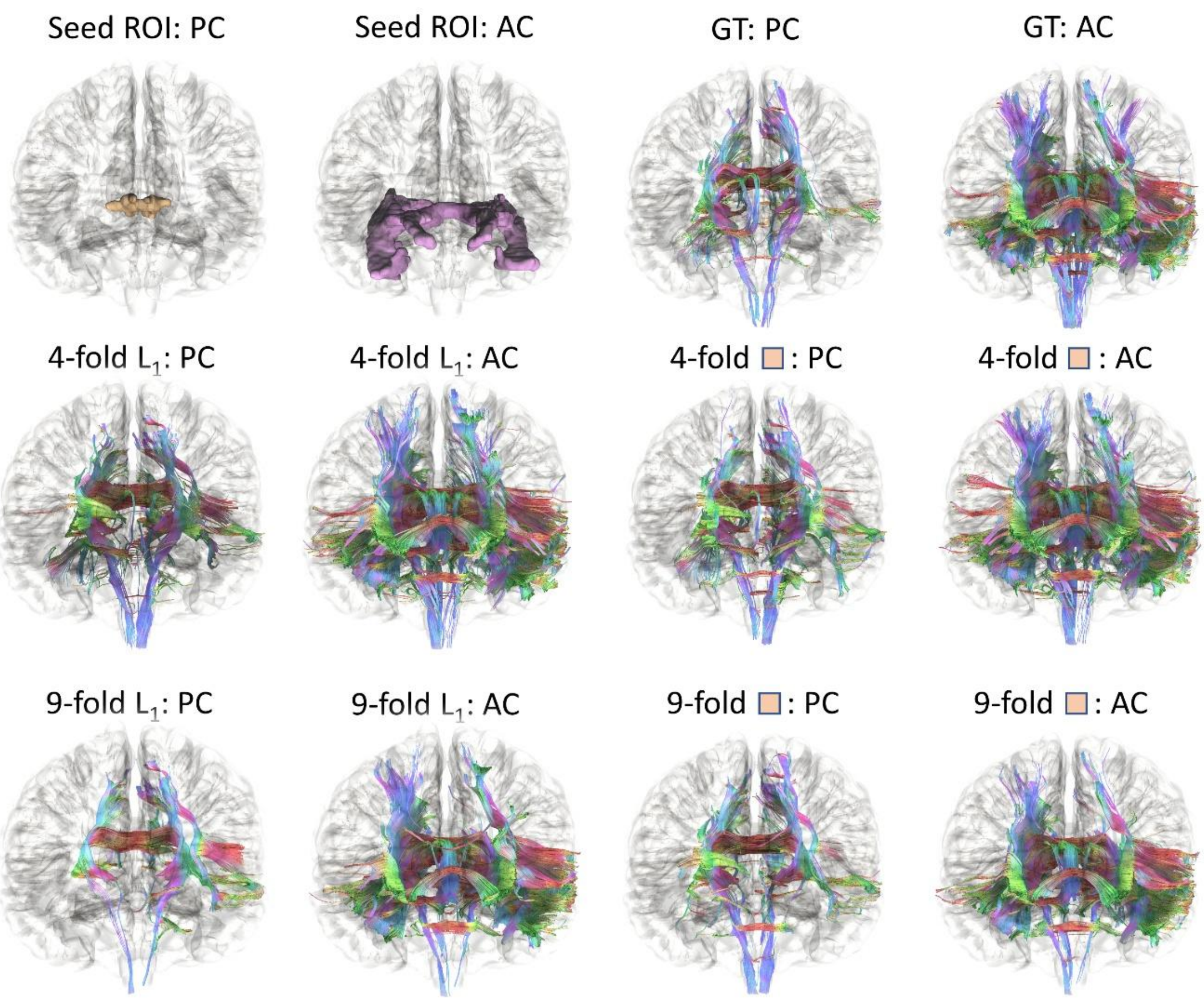

***Figure 4: Tractography results for 4-fold and 9-fold resolution enhancement for one of the volunteers in the test set.*** *The top row depicts seeding regions of posterior and anterior commissure (PC and AC) and tractography derived from the ground truth HCP data. The center row shows 4-fold upsampled data using the baseline model ($L_1$ loss) as well as the model based on the shallowest layer (red box). The bottom row shows 9-fold upsampled data using the same two models. Tractography parameters were quantitative anisotropy (QA) threshold: 0.1, angular threshold: 60°, step size: 0.5 mm, minimal tract length: 30 mm, maximal tract length: 200 mm, seeds: 1000000. 200000 tracts are visualized. Regions (PC and AC) were selected separately for seeding.*

Ablation and isolation studies provided details on the impact of the various VGG16 layers on model performance, revealing the link between layer depth and artifact burden. The studies further demonstrated that a loss based on solely the shallowest layer did not cause any artifacts, independent of the SNR of the input image. Shallow layers are thus preferable for feature-based losses for resolution enhancement of MR-diffusion data, when using a VGG16. Some of the degradation in image quality scores (PSNR, MAE, SSIM) is expected, when adding a loss term that is not pixel-based. Feature-loss related artifacts have also been reported in the context of edge enhancement[23] and phase retrieval[24] in natural images as well as deblurring of microscopy images[25]. The latter two studies found the shallowest layer to provide the best results and that image degradations like noise or blurring aggravated resulting artifact burden in other cases. Deng et al.[24] further showed that the grid-like artifact has a distinct frequency signature and indicate a connection between the artifacts and maxpooling-layers of the VGG16. These reports are in line with our observations.

While images derived from the baseline and the model using the shallowest layer were visually quite similar, derived diffusion parameters in various brain regions and tractography differed. On average, the feature-based loss led to higher QA values compared to the ground truth, while the baseline appeared to underestimate QA. Improved tractography results for the feature-based model as well as the depiction of additional tracts may directly be linked to the higher QA, since this parameter was used as a termination criterium during tracking. Similar offsets between feature-based models and the baseline were found for diffusion tensor parameters (FA, RD) and generalized q-space imaging parameters (RDI, ISO). In combination with the tracking results, this suggests that the various diffusion gradients may indeed be reflected in features of varying complexity and that diffusion effects may be learned indirectly via feature-based losses. While beyond the scope of this paper, future work may identify and eliminate the artifact origin for deeper layers and combinations thereof to further corroborate this and fully utilize feature-based losses.

All feature-based models as well as the baseline improved diffusion correlation in the test set. This indicates that deeper features contributed meaningful information, somewhat balancing the increased artifact burden. Since all models contained a pixel-based loss, it is expected that no distinct changes in the diffusion contrast were observed. Feature or perception losses have been increasingly applied in MRI[17,18,21,26-28] without reporting loss-related artifacts. There are three main reasons for this difference in observation: 1) Images exhibited distinctly higher SNR compared to the high-resolution $b_{1000}$ and $b_{2000}$ diffusion-weighted images in this study. If we were solely considering images without diffusion weighting, we would very likely have reached a similar conclusion. 2) Artifacts were not considered as such or not registered as a product of the loss function. 3) Most studies added additional loss terms like generative adversarial networks or regularization like total variation. In these cases, potential artifacts may be masked and the contributions from the individual loss terms are unclear, and a dedicated ablation study would be required to disentangle them. Accumulated data suggests that the full potential of feature-based losses may not be accessible in low SNR conditions when using feature extractors containing maxpooling-layers. Finding an architecture that enables use of deeper layers to generate artifact free images may further improve diffusion correlation and preserve high frequency features, while also addressing the smoothing observed when using solely shallow layers. Until then, we recommend that studies aiming to apply feature-based losses should evaluate its effect in isolation, particularly in low SNR regimes.

We did observe a decrease in performance for images exhibiting increasing diffusion weighting, despite normalization to address the varying signal intensities between the different diffusion weightings. This means that the models perform worse on diffusion data with high diffusion weighting. Most likely this is caused by low signal intensities in highly diffusion-weighted images which leads to smaller losses being propagated during the training. Future studies could alleviate this by either increasing the number of epochs to reach regimes where the loss contributions of all DWIs are similar or by adding a weight factor to the loss calculation which balances the differences in mean intensities in $b_0$, $b_{1000}$, and $b_{2000}$ images.

We found diffusion MRI to be beneficial for SR deep learning applications since datasets contain scans for the various gradient directions. Every scan therefore carries a multiple of images compared to standard anatomical scans. This allowed us to train a small memory footprint UNet (ResNet18 encoder) for 4-fold and 9-fold SR of diffusion MRI based on solely n=10 datasets. While the small-scale model enhanced diffusion quality for 4-fold resolution enhancement, the feasibility test for 9-fold enhancement indicated that restoration of details from further degraded data (e.g., 16-fold) will be severely restricted. Improved model performance in these cases may be achieved using deeper models or more modern CNN- (e.g. convnext[29] or regnet[30]) or transformer-based (e.g. dual attention vision transformer[31]) encoder architectures.

Despite the number of images used in this study being quite high, the number of datasets used to train our models (n=10) was rather low and a conscious compromise regarding time and computation constraints. While beyond the scope of this paper, the young adult HCP dataset contains 170 datasets, where both 7T and 3T diffusion scans are available, providing ample opportunity to scale the training and testing regimen to improve generalization or enable the training of deeper models.

## Methods

### Data, software, hardware

7T diffusion MRI data (n=15 volunteers) were collected from the human connectome project.[32] Details are listed in the supplementary information (Table S1). For readability, we use the terms $b_0$-, $b_{1000}$, and $b_{2000}$ image to refer to images acquired with b-values of 0, 1000, and 2000 s/mm$^2$. Each scan consists of 173 transversal slices of isotropic 1 mm resolution with 15 $b_0$, 64 $b_{1000}$, and 64 $b_{2000}$ images. All DWIs within a single scan were normalized together by dividing by $2^{16}$ to maintain relative intensities. Maximum intensities for each volunteer were saved to enable restoration to the original dynamic range. Pairs of axial low- and high-resolution images were generated via bicubic down sampling by factors of two and three for both axes, resulting in 4-fold and 9-fold lower total resolution (Figure 5). Models were trained to restore the full resolution from individual 2D LR input DWIs. Training, validation, and testing of all models was implemented using Pytorch (v2.3.1) and fastai (v2.7.15)[33]. Diffusion parameters based on GQI reconstruction as well as tractography were generated using DSI studio.[22,34] Model training and validation were run on a dedicated server using a NVIDIA A40 GPU (48GB), while inference for testing was performed on a local PC using a NVIDIA GeForce RTX 4060 Ti GPU (16GB).

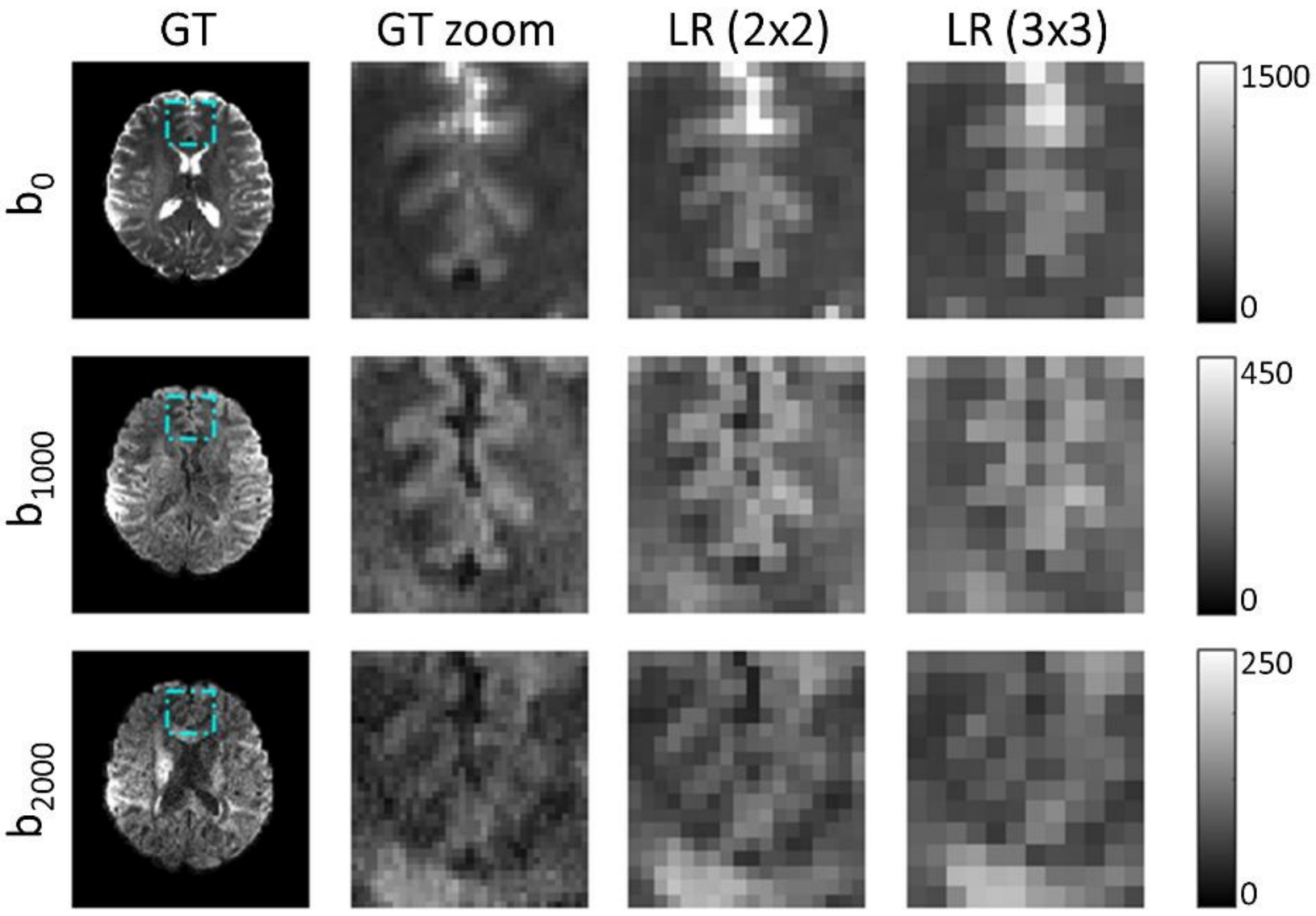

***Figure 5: Human connectome project (HCP) data.*** *The first column depicts a central transversal slice without ($b_0$) and with diffusion weighting ($b_{1000}$ and $b_{2000}$). These images are used as the ground truth (GT). The different diffusion weightings lead to different pixel intensity distributions. The cyan squares indicate the cropped area shown in the second to last column. Respective sections are downsampled by factor two (LR (2x2)) and three (LR (3x3)), illustrating the various resolutions employed in this study.*

**Model**

For model training, validation, and testing, n=197912, n=49478, and n=123695 images were used, which corresponds to an 8:2:5 split on case level. The split for training and validation was performed via random selection. Our SR model consisted of a pre-trained UNet[35] with a ResNet18[36] encoder to achieve a small memory footprint and fast training. We further applied weight decay and self-attention[37] to improve convergence and pixel shuffle initialized to convolution nearest-neighbour resize, which eliminates checkerboard artifacts typically present in randomly initialized sub-pixel convolutions.[38] Optimal learning rates were derived from the fastai[33] learning rate finder. Models were fine-tuned for 5 epochs using frozen encoder weights and 100 epochs using unfrozen weights.

**Loss terms**

Our baseline model in this study applied a pixel-wise loss $L_P$ based on the MAE

$$L_P = ||\hat{y} - y||_1, \quad (1)$$

where $\hat{y}$ is the image predicted by the model and $y$ the ground truth. For our feature-based loss we, settled on a VGG16 model (pretrained using ImageNet data) as a generator as described by Johnson et al.[16] During training, model predictions and ground truth were passed through the VGG16. Resulting feature maps at various intermediate layers $i$ were used for the loss as illustrated in Figure 6. We solely used activation maps prior to maxpooling-layers, which provided five blocks whose outputs $B_i(\hat{y})$ and $B_i(y)$ contributed to the feature loss $L_{F_i}$ at layer $i$ as follows:

$$L_{F_i} = \frac{1}{hwc} ||B_i(\hat{y}) - B_i(y)||_1, \quad (2)$$

where $h$, $w$, and $c$ are the height, width, and the number of channels of feature maps in layer $i$. The full feature loss $L_F$ corresponded to a weighted sum of the $n$ blocks involved in the experiment

$$L_F = \sum_{i=1}^{n} w_i L_{F_i} \quad (3)$$

where $w_i$ is a weighting factor used to match loss contributions of the individual blocks to $L_P$. We achieved this by performing a 1-epoch training for all our models, after which we determined $L_P$ and $L_{F_i}$. Weights $w_i$ were then calculated that $w_i L_{F_i} = L_p$. To anchor our final loss for our SR model to some image space information during training, we combined the feature loss and pixel-based loss terms:

$$L = L_P + L_F. \quad (4)$$

**Experiments**

Figure 7 illustrates the experiments and the evaluation of the trained models. To optimize a feature-based loss for SR of DWIs, two experiments were performed to identify how layer combinations and individual layers affected model performance compared to the baseline model. Models were trained for 4-fold resolution enhancement using diffusion data with respective downsampling for training. A third experiment assessed feasibility of a higher restoration factor (9-fold).

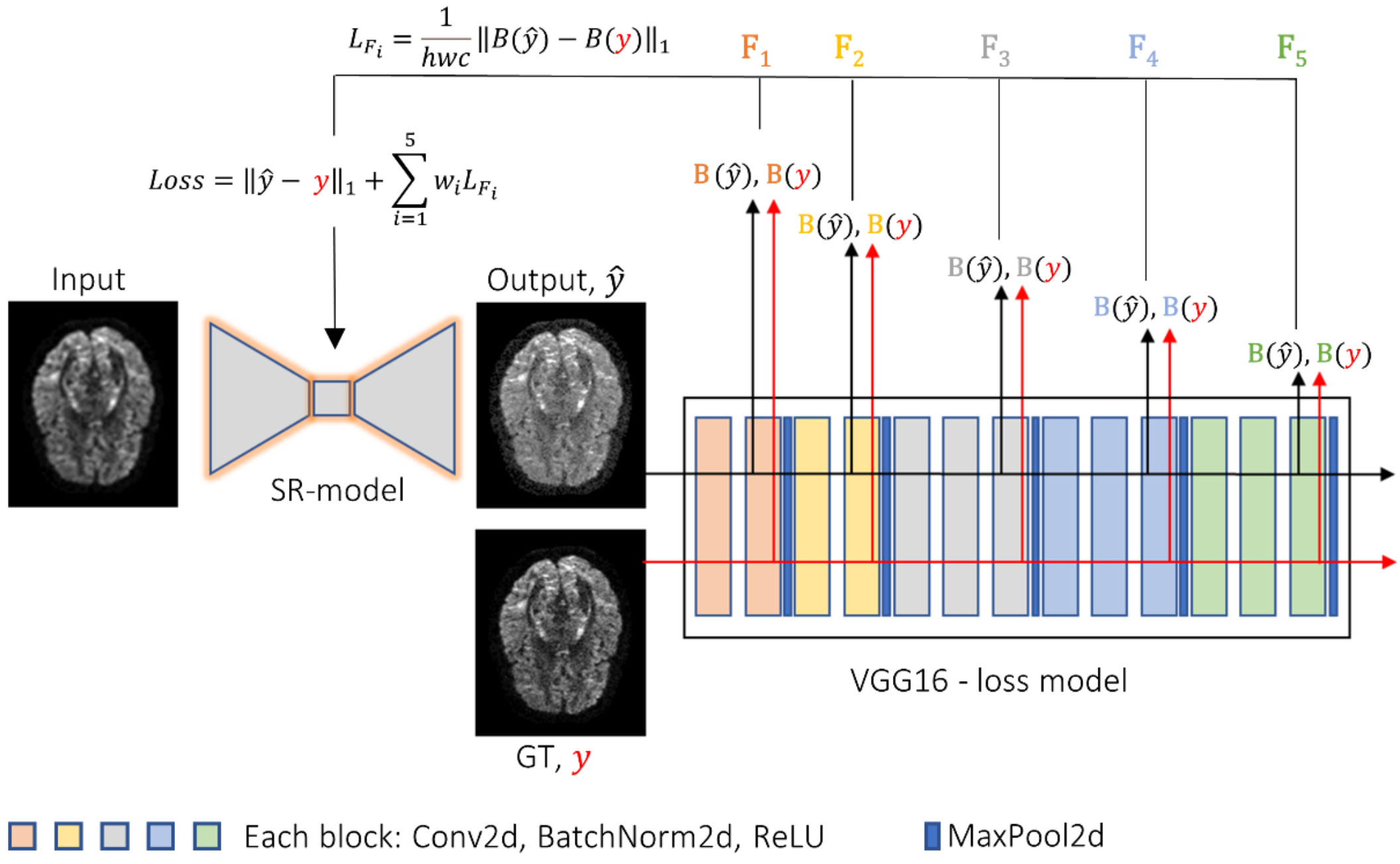

***Figure 6: Illustration of implemented feature loss.*** *Feature-based losses for the SR model were generated using inference of a VGG16 for SR-model output $\hat{y}$ and ground truth y. $L_1$-losses were calculated using the features from five layers located prior to the maxpooling layers. Each loss term was scaled to match the standard pixel-wise $L_1$-loss using a weight $w_i$. The SR-model consisted of a UNet with a ResNet18 encoder.*

**Experiment 1: feature ablation study** aimed to assess the impact of layer combinations. Based on the expectations that all layers would contribute meaningful information and therefore improve overall image quality, we started out with a model that combined feature maps from all five blocks. In each step of the ablation, we trained a new model where the deepest block was dropped from contributing to the loss.

**Experiment 2: feature isolation study** aimed to isolate the impact of individual layers by training multiple models using solely the features from a single block.

**Experiment 3: resolution study** aimed to assess the optimized feature-based loss for the more difficult task of 9-fold resolution restoration. Here, optimized means, the best-performing model with respect to image quality and diffusion parameters from the prior two experiments was selected. This model and the baseline were fine-tuned (frozen weights: 5 epochs and unfrozen weights: 20 epochs) using the training data dedicated to 9-fold resolution restoration. Model performance was evaluated based on tractography quality to showcase feasibility and potential limitations.

**Evaluation**

Model performance for the ablation and isolation experiments was evaluated using 1) image quality scores PSNR, SSIM, and MSE; 2) DWI quality control scores[39] (diffusion contrast and correlation); and 3) atlas-based (HCP842) comparison of diffusion parameters. For image quality score calculation, the image intensity values of individual DWIs were re-scaled to [0, 1] to account for diffusion weighting-induced differences in signal intensities between images. For DWI quality control scores and processing in DSI Studio, generated SR DWIs were scaled to their original dynamic range.

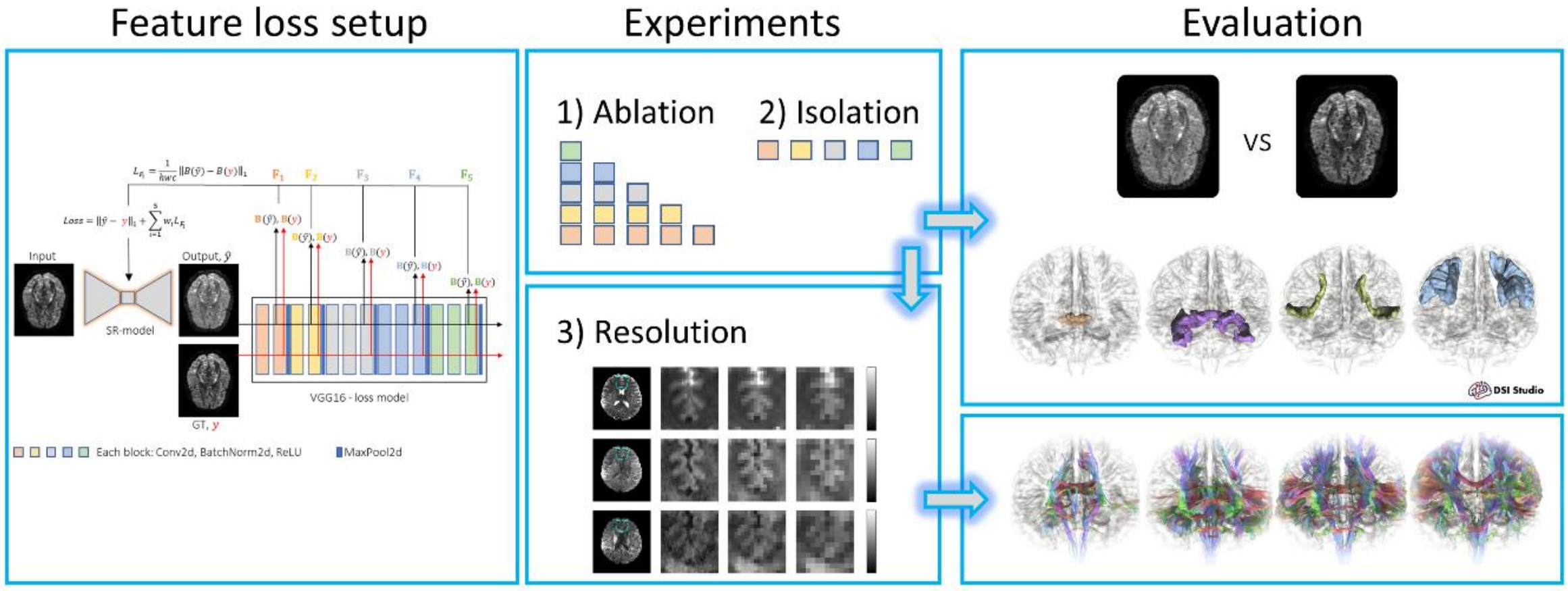

***Figure 7: Study overview.*** *We performed three separate experiments using feature-based losses derived from five distinct layers of a pre-trained VGG16 model. Coloured boxes indicate the layers that contributed to the loss. 1) We started with a model that used all five layers and then moved to train further models, where in each step of the ablation, the deepest layer was dropped from contributing to the loss. 2) We trained five models, where the feature-based loss was derived from the five individual layers. Models in these two experiments were trained for 4-fold resolution restoration and were evaluated using image similarity scores: absolute error (MAE), mean squared error (MSE), peak signal-to-noise ratio (PSNR), and SSIM (structural similarity index measure) as well as diffusion parameters derived in regions of interests (ROIs) of the HCP842 atlas. 3) The best-performing model of the first two experiments and the baseline were fine-tuned for 9-fold resolution restoration to assess feasibility and limitations. Both models were evaluated based on tractography quality.*

**Image quality scores**

Considering the image pair $(\hat{y}, y)$, PSNR is defined as

$$PSNR = 10log_{10}\frac{R(y)^2}{MSE(\hat{y}, y)} \quad 5$$

where, R(y) denotes the range of values inherent to the GT image and MSE$(\hat{y}, y)$ the mean squared error

$$MSE = \frac{1}{n}\sum_{i=1}^{n}(y_i - \hat{y}_i)^2. \quad 6$$

with $y_i$ and $\hat{y}_i$ being intensity values at pixel $i$ from $n$ total pixels. Higher values for PSNR are associated with improved image quality as are lower values for MSE. While MSE and PSNR quantify pixel-based errors and thus offer no information on pixel relationships and human visual perception, SSIM was designed as an image quality score providing feedback regarding the perceived quality of an image via luminance, contrast, and structure. For the image pair $(\hat{y}, y)$, SSIM is calculated as

$$SSIM = \frac{(2\mu_{\hat{y}}\mu_y + c_1)(2\sigma_{\hat{y}y} + c_2)}{(\mu_{\hat{y}}^2 + \mu_{\hat{y}}^2 + c_1)(\sigma_{\hat{y}}^2 + \sigma_{\hat{y}}^2 + c_2)} \quad 7$$

where $\mu_{\hat{y}}$ and $\mu_y$ denote the pixel sample means of $\hat{y}$ and y, $\sigma_{\hat{y}}^2$ and $\sigma_{\hat{y}}^2$ the variance of $\hat{y}$ and $y$, and $\sigma_{\hat{y}y}$ their covariance. The entries $c_1$ and $c_2$ denote stability constants defined as

$$c_1 = (k_1R)^2 \quad 8$$

and

$$c_2 = (k_2R)^2 \quad 9$$

where $R = R(y)$ again denotes the dynamic range of the pixel-values and $k_1$, $k_2$ tuneable values <<1. Images were masked using the brain mask provided as part of the HCP dataset prior to image quality assessment, meaning that only voxels containing brain tissue contributed to PSNR, MSE, and SSIM values.

**Quality control of diffusion data**

Quality control scores were generated using DSI studio.[39] The first quality score corresponded to the diffusion contrast, which estimates the contrast in DWIs generated by the diffusion gradient. A higher value is associated with increased likelihood to resolve fiber bundle orientations. The second score was the mean Pearson correlation coefficient of the neighbouring diffusion-weighted images:

$$\frac{1}{n}\sum_{i=1}^{n} \rho\left(S_i, S_{N(i)}\right), \tag{10}$$

where ρ calculates the Pearson correlation coefficient and $S_i$ is the i-th DWI. For each slice, there exists a series of $i$ DWIs, whose sequence is determined by the diffusion gradient table that was applied during the image acquisition. The index $N(i)$ for the neighbouring diffusion-weighted image $S_{N(i)}$ was given by the most similar diffusion sensitization expressed using diffusion gradient direction and weighting

$$N(i) = argmin_{j,j\neq i}\sqrt{b(i)}\vec{g}(i) - \sqrt{b(j)}\vec{g}(j), \tag{11}$$

where $b(i)$ is the b-value and $\vec{g}(i)$ the gradient direction of the i-th DWI. Higher correlation values indicate higher diffusion coherence within the dataset. In a typical measurement, decreases in this score suggest prominent eddy current artifacts, head motion, or any head coil issues that impact the diffusion signals. In this study, they are associated with SR restoration errors.

**Diffusion parameters**

Since we aimed to assess the performance of the SR model in a way that involves data of all b-values and thus images with significantly different SNR, we settled on GQI, which calculates the empirical distribution of water diffusion directly from the MR signal of the diffusion-weighted images without any assumption of the underlying distribution (e.g., Gaussian distribution DTI) of the molecule displacement. Rather than quantifying how fast the diffusion is, like in diffusion tensor imaging (DTI), this empirical distribution, named spin distribution function, quantifies the accumulated spin density of restricted diffusion sampled at any orientations. Contrary to DTI this allows resolving multiple fibre bundle orientations within a single voxel and the differentiation of restricted and unrestricted diffusion by utilizing data from all shells. GQI requires the selection of the parameter σ, the diffusion sampling ratio, which controls the displacement range of the diffusing spins. In this study, a value of 1.75 was empirically found to provide good quality in regions of crossing and non-crossing fibre bundles.

Derived parameters following GQI reconstruction were QA, RDI, and ISO. For reference, we also tracked the DTI parameters FA and RD. To assess SR prediction quality in all parts of the brain in a physiologically meaningful way we collected statistics for these parameters in 80 anatomical ROI using the HCP842[40] atlas provided in DSI studio. Diffusion data was first co-registered with the template space of the atlas to facilitate alignment. Since DSI Studio could not perform co-registration to this template space for downsampled data, it was omitted from this analysis. To enable comparison of all ROIs we used the difference between SR data and the ground truth. In addition to statistics averaged over all HCP842 ROIs we selected statistics for the SLF and MLF as well as the AC to illustrate various areas of the brain.

**Tractography**

While many pathological changes can already be assessed based on pixel-based diffusion parameters, there are downstream analyses which go beyond that. So-called fibre tracking uses the resolved diffusion orientations to track the diffusion process throughout the brain, enabling the direct assessment of brain connectivity. Knowledge about fibre tract locations associated with specific tasks (e.g., memory, motor skills, etc.) may improve tumour resection or radiation planning or the assessment of connectivity under pathologic conditions. For the resolution study we therefore performed tractography using a restricted number of brain regions (AC, PC, SLF, MLF) as seed points for the two levels of resolution restoration (4-fold, 9-fold) as well as the ground truth. Since ROI-based tracking was not possible for downsampled data, we used our baseline model for reference. Tractography parameters were QA threshold: 0.1, angular threshold: 60°, step size: 0.5 mm, minimal tract length: 30 mm, maximal tract length: 200 mm, seeds: 1000000. QA, RDI, and FA were quantified within the tracts and tract quality was assessed visually to assess potential limitations of the trained models.

**Statistics**

All data was analyzed for normal distribution using a Shapiro-Wilk test with $p < 0.05$. Subsequent tests for significant differences ($p < 0.05$) were performed using a paired t-test in case of normality (otherwise Wilcoxon-rank sum test). Bonferroni corrections were applied in cases of multiple tests.

**Resource availability**

***Lead contact***

Requests for further information and resources should be directed to and will be fulfilled by the lead contact, David Lohr (d.lohr@uke.com).

***Materials availability***

This study did not generate new unique reagents or materials.

***Data and code availability***

Human connectome data is freely available for research purposes and can be accessed here: https://balsa.wustl.edu/project?project=HCP_YA. The study numbers for the data used in this specific study are listed in the supplementary information. All original code (data processing, model training and inference) has been deposited at https://github.com/IPMI-ICNS-UKE/Sharp-diffusion .

**Acknowledgements**

Funded by the Deutsche Forschungsgemeinschaft (DFG, German Research Foundation) - 519067094.

We thank Sara Tiedemann for editorial support.

**Author contributions**

**DL** led conceptualization, project administration and acquired funding. He further performed data curation, formal analysis, wrote the code, performed validation, created figures for visualization and wrote the original draft. **RW** provided computing resources, review input of the draft, figures, tables, and supplementary materials, refinement regarding language and suggestions regarding content.

**Declaration of interests**

There are no conflicts of interest to declare.

**Supplemental information**

Figures S1-S7. Table S1.

**References**

1. Salat, D.H. (2014). Chapter 12 - Diffusion Tensor Imaging in the Study of Aging and Age-Associated Neural Disease. In Diffusion MRI (Second Edition), H. Johansen-Berg, and T.E.J. Behrens, eds. (Academic Press), pp. 257–281. 10.1016/B978-0-12-396460-1.00012-3.
2. Goveas, J., O'Dwyer, L., Mascalchi, M., Cosottini, M., Diciotti, S., De Santis, S., Passamonti, L., Tessa, C., Toschi, N., and Giannelli, M. (2015). Diffusion-MRI in neurodegenerative disorders. Magnetic resonance imaging *33*, 853–876. 10.1016/j.mri.2015.04.006.
3. Tournier, J.D. (2019). Diffusion MRI in the brain - Theory and concepts. Prog Nucl Magn Reson Spectrosc *112-113*, 1–16. 10.1016/j.pnmrs.2019.03.001.
4. Lerch, J.P., van der Kouwe, A.J.W., Raznahan, A., Paus, T., Johansen-Berg, H., Miller, K.L., Smith, S.M., Fischl, B., and Sotiropoulos, S.N. (2017). Studying neuroanatomy using MRI. Nature Neuroscience *20*, 314–326. 10.1038/nn.4501.
5. Van Essen, D.C., Smith, S.M., Barch, D.M., Behrens, T.E.J., Yacoub, E., and Ugurbil, K. (2013). The WU-Minn Human Connectome Project: An overview. Mapping the Connectome *80*, 62–79. 10.1016/j.neuroimage.2013.05.041.
6. Alexander, D.C., Zikic, D., Ghosh, A., Tanno, R., Wottschel, V., Zhang, J., Kaden, E., Dyrby, T.B., Sotiropoulos, S.N., Zhang, H., and Criminisi, A. (2017). Image quality transfer and applications in diffusion MRI. NeuroImage *152*, 283–298. 10.1016/j.neuroimage.2017.02.089.
7. Elsaid, N.M.H., and Wu, Y.C. (2019). Super-Resolution Diffusion Tensor Imaging using SRCNN: A Feasibility Study. 23–27 July 2019. pp. 2830–2834.
8. Tian, Q., Li, Z., Fan, Q., Ngamsombat, C., Hu, Y., Liao, C., Wang, F., Setsompop, K., Polimeni, J.R., Bilgic, B., and Huang, S.Y. (2021). SRDTI: Deep learning-based super-resolution for diffusion tensor MRI. Preprint at arXiv. 10.48550/arXiv.2102.09069.
9. Lyon, M., Armitage, P., and Álvarez, M.A. (2022). Angular Super-Resolution in Diffusion MRI with a 3D Recurrent Convolutional Autoencoder. Preprint at arXiv. 10.48550/arXiv.2203.15598.
10. Chen, G., Hong, Y., Huynh, K.M., and Yap, P.T. (2023). Deep learning prediction of diffusion MRI data with microstructure-sensitive loss functions. Med Image Anal *85*, 102742. 10.1016/j.media.2023.102742.
11. Ren, M., Kim, H., Dey, N., and Gerig, G. (2021). Q-space Conditioned Translation Networks for Directional Synthesis of Diffusion Weighted Images from Multi-modal Structural MRI. Medical image computing and computer-assisted intervention : MICCAI ... International Conference on Medical Image Computing and Computer-Assisted Intervention *12907*, 530–540. 10.1007/978-3-030-87234-2_50.
12. Qin, Y., Liu, Z., Liu, C., Li, Y., Zeng, X., and Ye, C. (2021). Super-Resolved q-Space deep learning with uncertainty quantification. Medical Image Analysis *67*, 101885. 10.1016/j.media.2020.101885.
13. Zhou, W., Bovik, A.C., Sheikh, H.R., and Simoncelli, E.P. (2004). Image quality assessment: from error visibility to structural similarity. IEEE Transactions on Image Processing *13*, 600–612. 10.1109/TIP.2003.819861.
14. Wang, Z., Simoncelli, E.P., and Bovik, A.C. (2003). Multiscale structural similarity for image quality assessment. 9–12 Nov. 2003. pp. 1398–1402 Vol.1392.

15. Zhang, L., Zhang, L., Mou, X., and Zhang, D. (2011). FSIM: A Feature Similarity Index for Image Quality Assessment. IEEE Transactions on Image Processing *20*, 2378–2386. 10.1109/TIP.2011.2109730.
16. Johnson, J., Alahi, A., and Fei-Fei, L. (2016). Perceptual Losses for Real-Time Style Transfer and Super-Resolution. Preprint at arXiv. 10.48550/arXiv.1603.08155.
17. Panda, A., Naskar, R., Rajbans, S., and Pal, S. (2019). A 3D Wide Residual Network with Perceptual Loss for Brain MRI Image Denoising. 6–8 July 2019. pp. 1–7.
18. Javadi, M., Sharma, R., Tsiamyrtzis, P., Shah, S., Leiss, E.L., and Tsekos, N.V. (2023). From Perception to Precision: Navigating Perceptual Loss in MRI Super-Resolution. 4–6 Dec. 2023. pp. 57–61.
19. Zhang, K., Hu, H., Philbrick, K., Conte, G.M., Sobek, J.D., Rouzrokh, P., and Erickson, B.J. (2022). SOUP-GAN: Super-Resolution MRI Using Generative Adversarial Networks. Tomography (Ann Arbor, Mich.) *8*, 905–919. 10.3390/tomography8020073.
20. Yang, G., Yu, S., Dong, H., Slabaugh, G., Dragotti, P.L., Ye, X., Liu, F., Arridge, S., Keegan, J., Guo, Y., and Firmin, D. (2018). DAGAN: Deep De-Aliasing Generative Adversarial Networks for Fast Compressed Sensing MRI Reconstruction. IEEE transactions on medical imaging *37*, 1310–1321. 10.1109/TMI.2017.2785879.
21. Lyu, Q., Shan, H., Steber, C., Helis, C., Whitlow, C., Chan, M., and Wang, G. (2020). Multi-Contrast Super-Resolution MRI Through a Progressive Network. IEEE transactions on medical imaging *39*, 2738–2749. 10.1109/TMI.2020.2974858.
22. Fang-Cheng, Y., Wedeen, V.J., and Tseng, W.Y.I. (2010). Generalized Q-Sampling Imaging. Medical Imaging, IEEE Transactions on *29*, 1626–1635. 10.1109/TMI.2010.2045126.
23. Fu, Z., Zheng, Y., Ma, T., Ye, H., Yang, J., and He, L. (2022). Edge-aware deep image deblurring. Neurocomputing *502*, 37–47. 10.1016/j.neucom.2022.06.051.
24. Deng, M., Goy, A., Li, S., Arthur, K., and Barbastathis, G. (2020). Probing shallower: perceptual loss trained Phase Extraction Neural Network (PLT-PhENN) for artifact-free reconstruction at low photon budget. Opt. Express *28*, 2511–2535. 10.1364/OE.381301.
25. Krawczyk, P., Gaertner, M., Jansche, A., Bernthaler, T., and Schneider, G. (2024). Reducing artifact generation when using perceptual loss for image deblurring of microscopy data for microstructure analysis. Methods in Microscopy *1*, 137–150. 10.1515/mim-2024-0012.
26. Ghodrati, V., Shao, J., Bydder, M., Zhou, Z., Yin, W., Nguyen, K.L., Yang, Y., and Hu, P. (2019). MR image reconstruction using deep learning: evaluation of network structure and loss functions. Quant Imaging Med Surg *9*, 1516–1527. 10.21037/qims.2019.08.10.
27. Sarasaen, C., Chatterjee, S., Breitkopf, M., Rose, G., Nürnberger, A., and Speck, O. (2021). Fine-tuning deep learning model parameters for improved super-resolution of dynamic MRI with prior-knowledge. Artificial intelligence in medicine *121*, 102196. 10.1016/j.artmed.2021.102196.
28. Jannat, S.R., Lynch, K., Fotouhi, M., Cen, S., Choupan, J., Sheikh-Bahaei, N., Pandey, G., and Varghese, B.A. (2025). Advancing 1.5T MR imaging: toward achieving 3T quality through deep learning super-resolution techniques. Frontiers in Human Neuroscience *19*. 10.3389/fnhum.2025.1532395.
29. Liu, Z., Mao, H., Wu, C.Y., Feichtenhofer, C., Darrell, T., and Xie, S. (2022). A ConvNet for the 2020s. 18–24 June 2022. pp. 11966–11976.
30. Radosavovic, I., Kosaraju, R.P., Girshick, R., He, K., and Dollár, P. (2020). Designing Network Design Spaces. 13–19 June 2020. pp. 10425–10433.
31. Ding, M., Xiao, B., Codella, N., Luo, P., Wang, J., and Yuan, L. (2022). DaViT: Dual Attention Vision Transformers. held in Cham, 2022. S. Avidan, G. Brostow, M. Cissé, G.M. Farinella, and T. Hassner, eds. (Springer Nature Switzerland), pp. 74–92.
32. Van Essen, D.C., Ugurbil, K., Auerbach, E., Barch, D., Behrens, T.E., Bucholz, R., Chang, A., Chen, L., Corbetta, M., Curtiss, S.W., et al. (2012). The Human Connectome Project: a data acquisition perspective. NeuroImage *62*, 2222–2231. 10.1016/j.neuroimage.2012.02.018.
33. Howard, J., and Gugger, S. (2020). Fastai: A Layered API for Deep Learning. Information *11*, 108.
34. Yeh, F.-C. DSI Studio. http://dsi-studio.labsolver.org. Accessed October 22, 2024.
35. Ronneberger, O., Fischer, P., and Brox, T. (2015). U-Net: Convolutional Networks for Biomedical Image Segmentation. Medical Image Computing and Computer-Assisted Intervention – MICCAI 2015, 2015. N. Navab, J. Hornegger, W.M. Wells, and A.F. Frangi, eds. (Springer International Publishing), pp. 234–241.
36. He, K., Zhang, X., Ren, S., and Sun, J. (2016). Deep Residual Learning for Image Recognition. held in Las Vegas, NV, June 01, 2016. pp. 1.
37. Zhang, H., Goodfellow, I., Metaxas, D., and Odena, A. (2018). Self-Attention Generative Adversarial Networks. Preprint at arXiv. 10.48550/arXiv.1805.08318.

38. Aitken, A., Ledig, C., Theis, L., Caballero, J., Wang, Z., and Shi, W. (2017). Checkerboard artifact free sub-pixel convolution: A note on sub-pixel convolution, resize convolution and convolution resize. Preprint at arXiv. 10.48550/arXiv.1707.02937.
39. Yeh, F.-C., Zaydan, I.M., Suski, V.R., Lacomis, D., Richardson, R.M., Maroon, J.C., and Barrios-Martinez, J. (2019). Differential tractography as a track-based biomarker for neuronal injury. NeuroImage *202*, 116131. 10.1016/j.neuroimage.2019.116131.
40. Yeh, F.-C., Panesar, S., Fernandes, D., Meola, A., Yoshino, M., Fernandez-Miranda, J.C., Vettel, J.M., and Verstynen, T. (2018). Population-averaged atlas of the macroscale human structural connectome and its network topology. NeuroImage *178*, 57–68. 10.1016/j.neuroimage.2018.05.027.

**Supplemental information: Layer Selection in Feature-Based Losses Affects Image Quality and Microstructural Consistency in Deep Learning Super-Resolution of Brain Diffusion MRI**

David Lohr[1,2,3,*], Rene Werner[1,2,3]

[1]Institue for Applied Medical Informatics, University Medical Center Hamburg-Eppendorf, Hamburg, 20246, Germany

[2]Institute of Computational Neuroscience, University Medical Center Hamburg-Eppendorf, 20246, Hamburg, Germany

[3]Center for Biomedical Artificial Intelligence (bAIome), University Medical Center Hamburg-Eppendorf, 20246, Hamburg, Germany

*Corresponding and lead author: d.lohr@uke.de

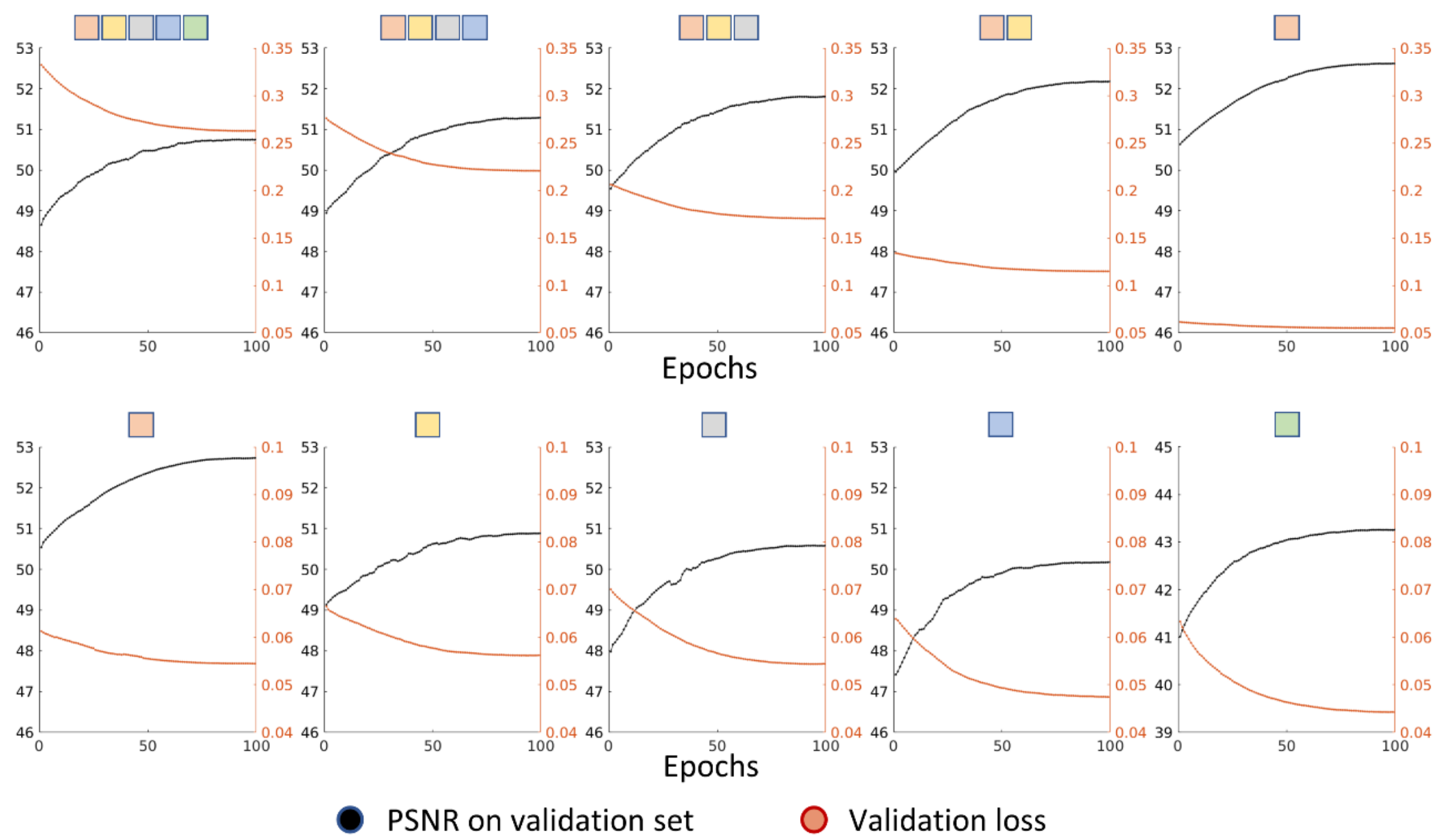

***Figure S1: Peak signal-to-noise ratio (PSNR) and loss development on the validation set during training of feature-based models.*** *The validation loss contained the sum of a pixel-based $L_1$ loss and various feature-based loss terms. Coloured boxes indicate the layers that contributed to the loss. Black curves denote PSNR values and red curves denote loss values. PSNR values are based on non-normalized images and thus differ from the assessment on the test in the manuscript. Combining deeper layers resulted in increasingly lower PSNR and higher loss on the validation set. The incremental increase in the loss was expected, since loss terms from an increasing number of channels were used. The validation loss for isolated models was comparable for all layers, but lowest for the deepest layer. PSNR values, however, went up for increasingly shallow layers. For visibility, the PSNR axis was adjusted for the isolated deepest layer (bottom right).*

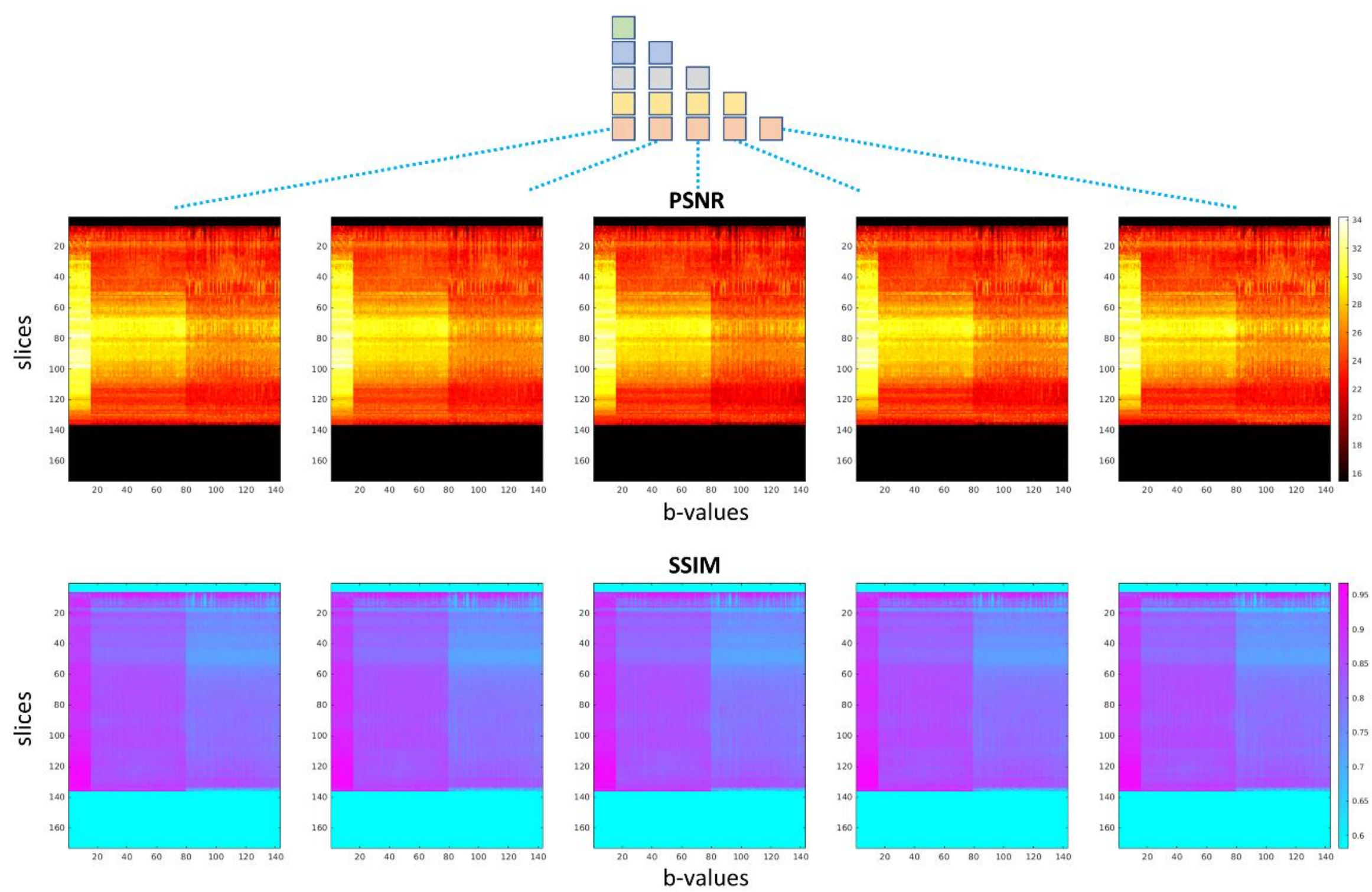

***Figure S2: Impact of slice position and b-value in the ablation study.*** *Depicted is data for a representative subject of the test set. Coloured boxes indicate the layers that contributed to the loss. Low slice numbers denote caudal slices while high slices numbers denote dorsal slices. Slices outside the brain mask were excluded from evaluation. Diffusion weightings were sorted in ascending order ($b_0$, $b_{1000}$, $b_{2000}$). Respective ranges are illustrated once in the top left but count for all graphics.* ***Top:*** *Peak signal-to-noise ratio (PSNR) distribution over slices and b-values. Black sections indicate slices outside the mask which were excluded from the evaluation.* ***Bottom:*** *Structural similarity index measure (SSIM) distribution over slices and b-values. Cyan sections indicate slices outside the mask which were excluded from the evaluation.*

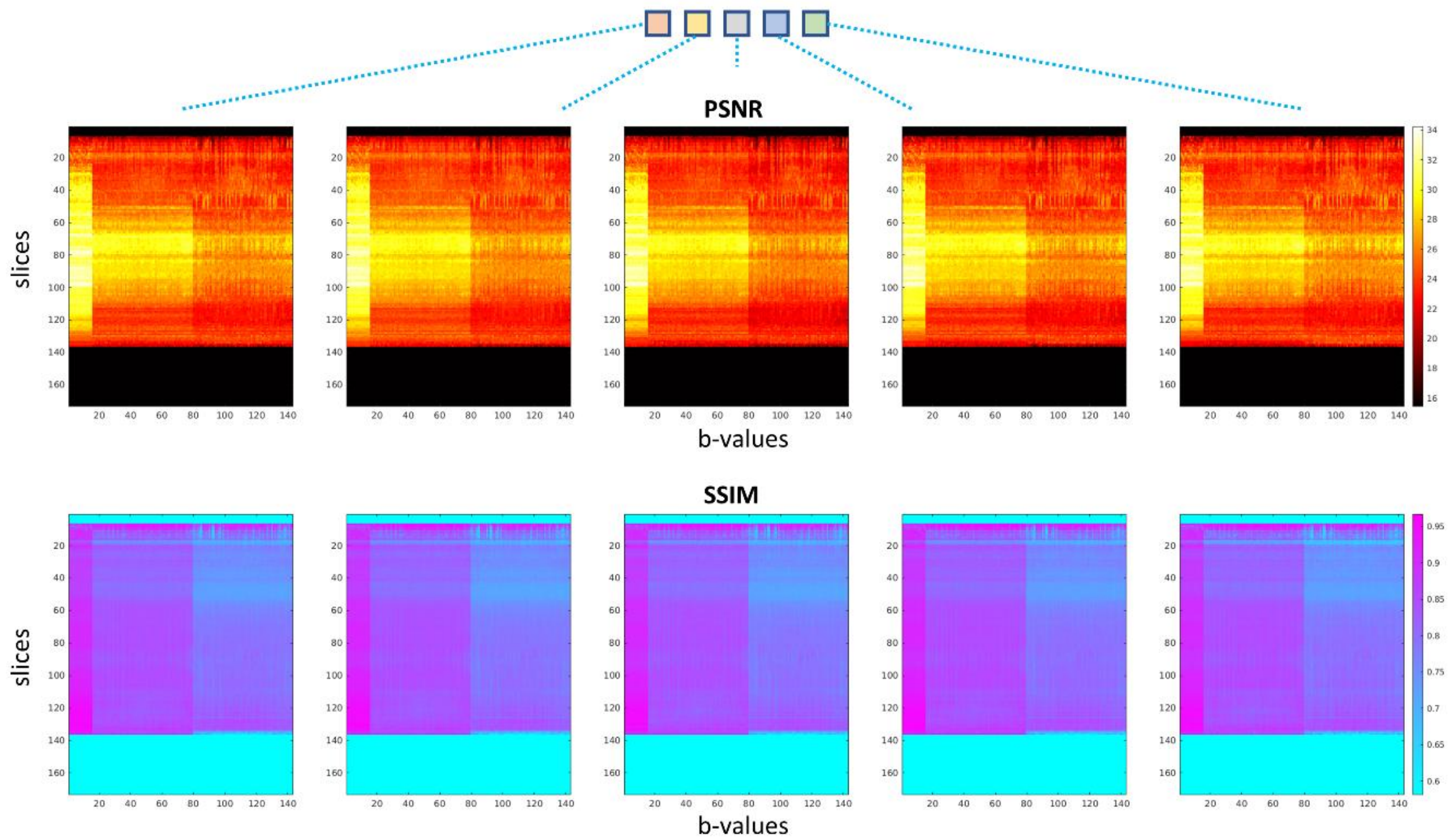

***Figure S3: Impact of slice position and b-value in the isolation study.*** *Depicted is data for a representative subject of the test set. Coloured boxes indicate the layers that contributed to the loss. Low slice numbers denote caudal slices while high slices numbers denote dorsal slices. Slices outside the brain mask were excluded from evaluation. Diffusion weightings were sorted in ascending order ($b_0$, $b_{1000}$, $b_{2000}$). Respective ranges are illustrated once in the top left but count for all graphics.* ***Top:*** *Peak signal-to-noise ratio (PSNR) distribution over slices and b-values. Black sections indicate slices outside the mask which were excluded from the evaluation.* ***Bottom:*** *Structural similarity index measure (SSIM) distribution over slices and b-values. Cyan sections indicate slices outside the mask which were excluded from the evaluation.*

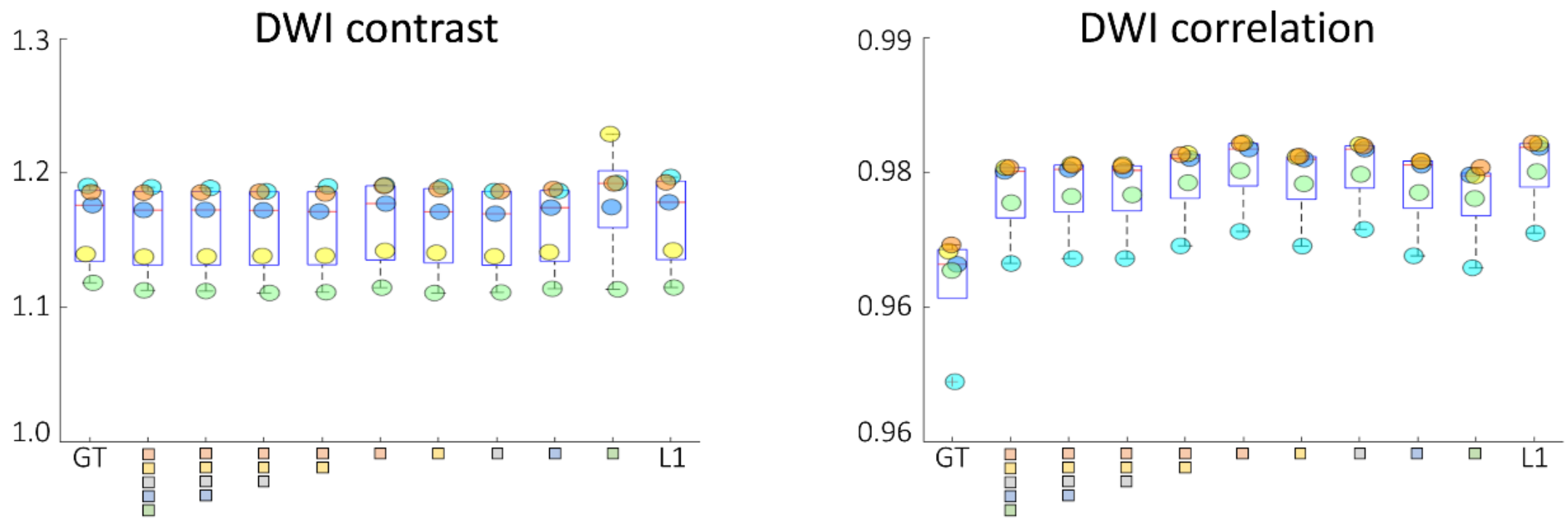

***Figure S4: Diffusion image quality scores in the test set. Left:*** *Diffusion weighted image (DWI) contrast. Feature-based models and the baseline showed no improvement in diffusion contrast compared to the ground truth.* ***Right:*** *DWI correlation. Feature-based models and the baseline showed distinct improvements in DWI correlation compared to the ground truth (GT). This indicates higher diffusion coherence within the resolution-enhanced diffusion data. The highest correlation was found for the model using solely the shallowest layer for the loss followed by the baseline. Colour coding indicates the five participants in the test set.*

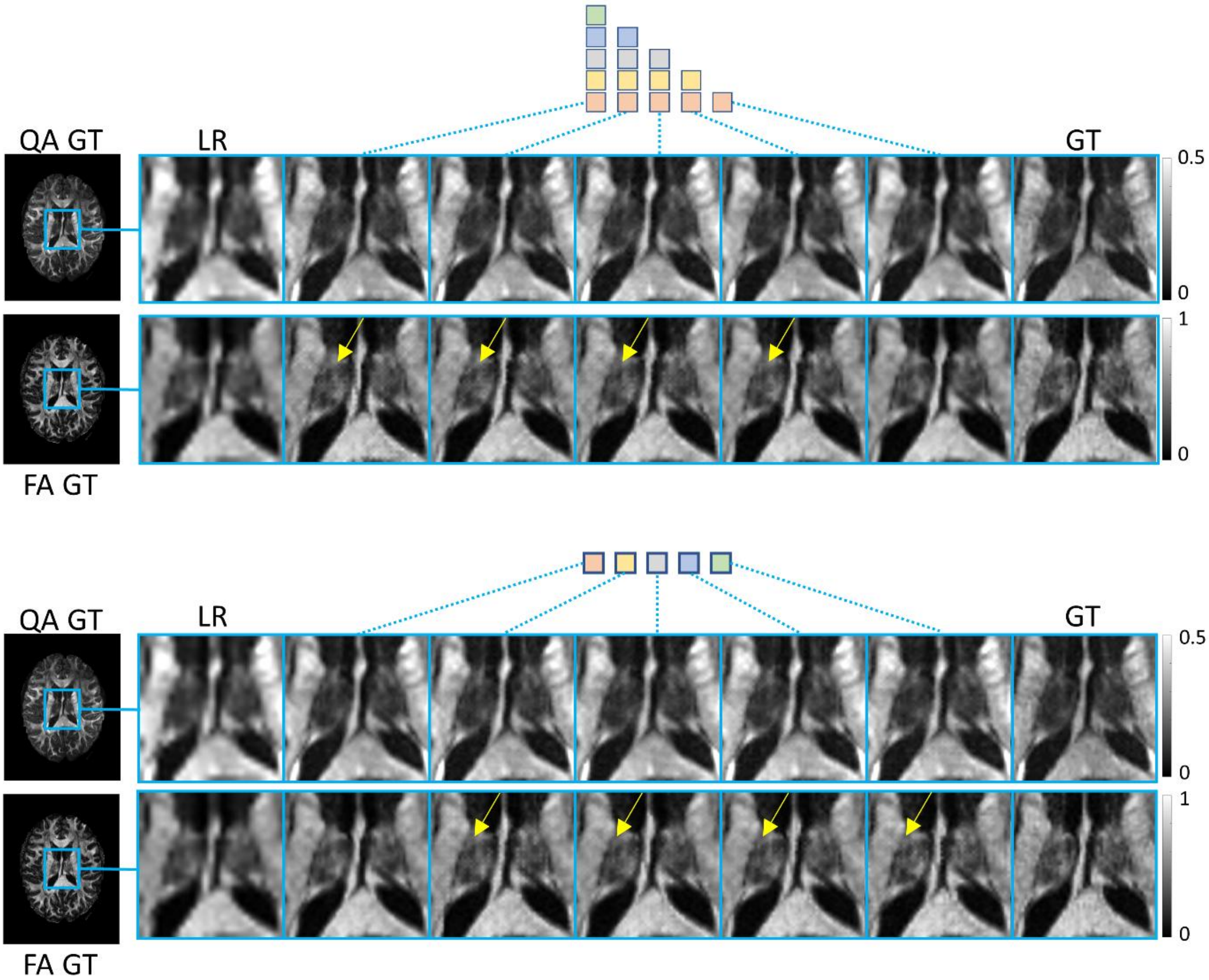

***Figure S5: Effects of feature-based losses on diffusion parameters QA and FA.*** *Ablation (top) and isolation (bottom) study results, showcasing improvements in resolution and detail compared to the low-resolution (LR) input. Coloured boxes indicate the layers that contributed to the loss. Cyan squares indicate crop sections employed for the low resolution (input), the various model predictions for that input, and the ground truth (GT). Yellow arrays indicate areas where grid-like artefacts previously observed in the raw diffusion weighted images are visible. While these are more pronounced in FA maps, they are also present for QA.*

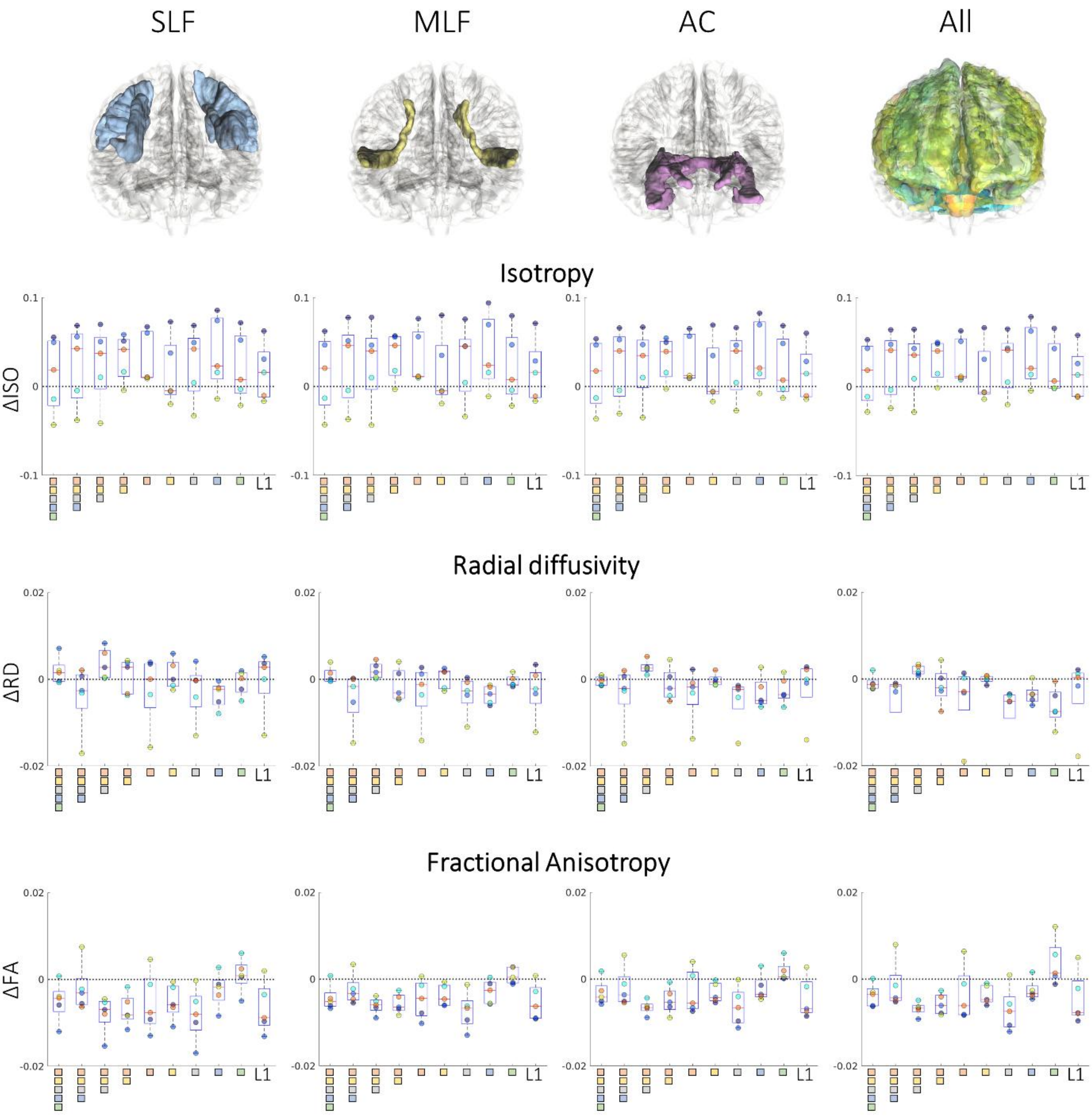

***Figure S6: Region-based diffusion parameters for models of the ablation and isolation study.** Regions of interest (ROI) were derived from the HCP842 atlas. **First row:** ROIs selected to illustrate model performance in various brain regions, namely superior longitudinal fasciculus (SLF), median longitudinal fasciculus (MLF), and anterior commissure (AC). The ROI "All" contains the full 80 ROIs within the atlas. **Second row:** Isotropy (ISO) difference between model predictions and ground truth on the test set. **Third row:** Difference in radial diffusivity (RD, diffusion tensor parameter) between model predictions and ground truth in the test set. **Fourth row:** Difference in fractional anisotropy (FA, diffusion tensor parameter) between model predictions and ground truth in the test set. Coloured boxes indicate the layers that contributed to the loss. Colour coding for the data points denotes different volunteers of the test set. The dotted line marks where prediction and ground truth would have been equal.*

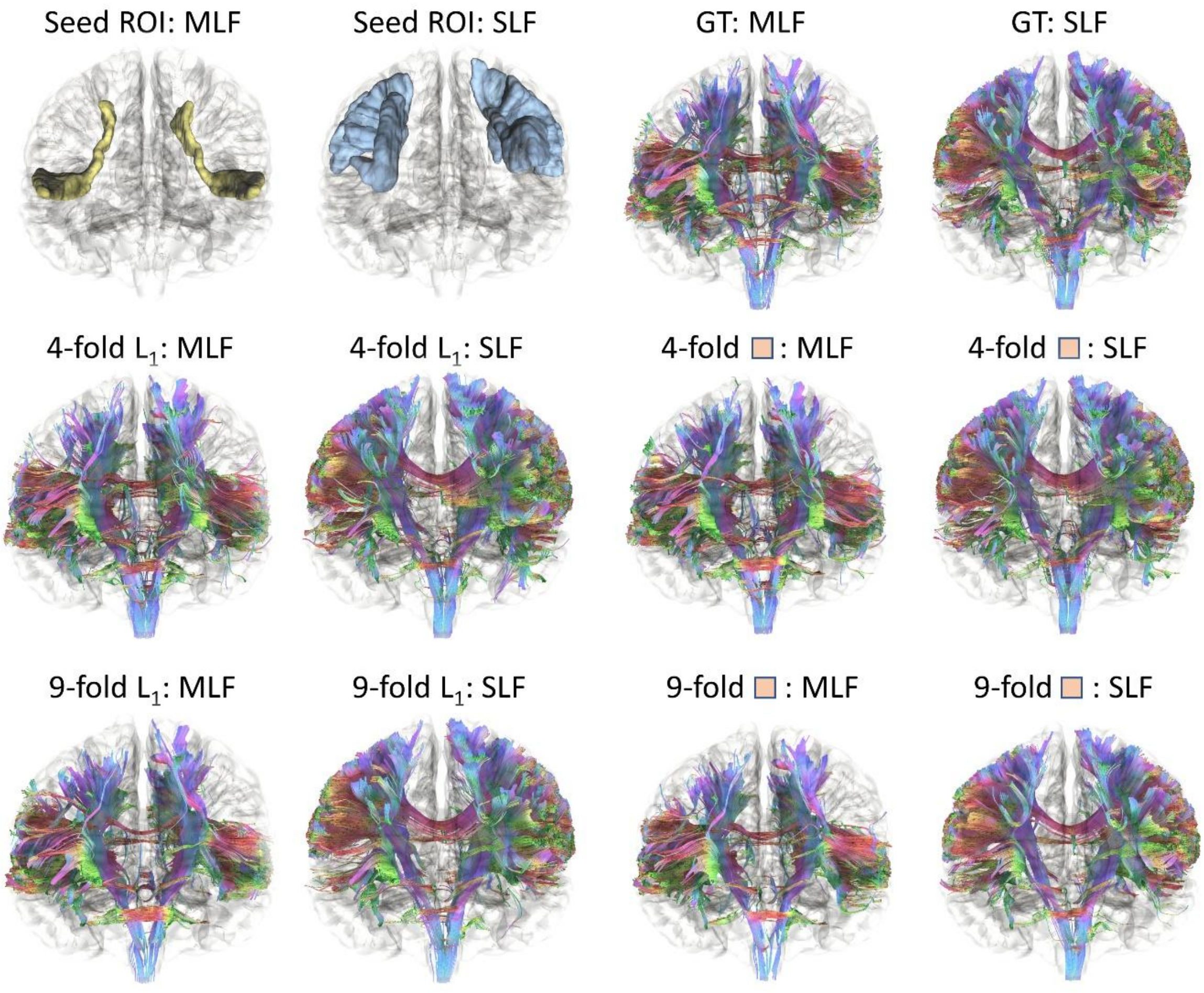

***Figure S7: Tractography results for 4-fold and 9-fold resolution enhancement for one of the volunteers in the test set.*** *The top row depicts seeding regions of medial and superior longitudinal fasciculus (MLF and SLF) and tractography derived from the ground truth HCP data. The center row shows 4-fold upsampled data using the baseline model ($L_1$ loss) as well as the model based on the shallowest layer (red box). The bottom row shows 9-fold upsampled data using the same two models. Tractography parameters were quantitative anisotropy (QA) threshold: 0.1, angular threshold: 60°, step size: 0.5 mm, minimal tract length: 30 mm, maximal tract length: 200 mm, seeds: 1000000. 200000 tracts are visualized. Regions (MLF and SLF) were selected separately for seeding.*

***Table S1: List of HCP identifiers for training and test sets.***

| | |
|---|---|
| **Training** | 100610, 102311, 102816, 104416, 105923, 108323, 109123, 111312, 111514, 114823 |
| **Testing** | 115017, 115825, 116726, 118225, 125525 |